# Crystallinity in Niobium oxides: A pathway for mitigation of Two-Level System Defects in Niobium 3D Resonator for quantum applications

Y. Kalboussi,[1] I. Curci,[1] F. Miserque,[2] D. Troadec,[3] N. Brun,[4] M. Walls,[4] G. Jullien,[1] F. Eozenou,[1] M. Baudrier,[1] L. Maurice,[1] Q. Bertrand,[1] P. Sahuquet[1] and T. Proslier[1]

[1]Institut des lois fondamentales de l'univers, Commissariat de l'énergie atomique-centre de saclay, Paris-Saclay university,91191 Gif sur Yvette, France.
[2]Université Paris-Saclay, Service de Recherche sur la Corrosion et le Comportement des Matériaux, 91191 Gif-sur-Yvette, France.
[3]Institut d'Electronique de Microélectronique et de Nanotechnologies, Université de Lille CNRS Université Polytechnique Hauts-de-France UMR 8520 – IEMN, Lille F-5900, France.
[4]Laboratoire de physique des solides, Paris-Saclay University, 91400 Orsay, France.

## ABSTRACT

Materials imperfections in Nb-based superconducting quantum circuits—in particular, two-level-system (TLS) defects—are a major source of decoherence, ultimately limiting the performance of quantum computation and sensing. Thus, identifying and understanding the microscopic origin of possible TLS defects in these devices will help developing strategies to eliminate them and is key to superconducting qubit performance improvement. In this paper, we demonstrate an order of magnitude reduction in two-level system losses in three-dimensional superconducting radio frequency (SRF) niobium resonators by a 10 hour high vacuum (HV) heat treatment at 650°C, even after exposure to air and high pressure rinsing (HPR). X-ray photoelectron spectroscopy (XPS) and high-resolution scanning transmission electron microscopy (STEM) reveal an alteration of the native oxide composition re-grown after air exposure and HPR and the creation of nano-scale crystalline oxide regions, which correlates with the measured tenfold quality factor enhancement at low fields of the 1.3 GHz niobium resonator. Tunneling spectroscopy measurements show a pronounced proximity effect that further confirms the presence of metallic layers on the niobium surface.

## I. INTRODUCTION

Superconducting niobium resonators, originally developed for particle acceleration with exceptionally high quality factors (Q > $10^{10}$-$10^{11}$), are gaining attention in quantum computing applications [1-5]. These resonators are excellent candidates for storing quantum information as quantum d-level systems (qudits) due to their remarkably long lifetimes of up to several seconds and their extensive accessible Hilbert spaces, which provide the potential for direct encoding of qudits and offer advantages over the two-level qubit encoding [3-5]. Niobium SRF cavities have been also proven very useful in studying the losses in two-dimensional superconducting qubits and measuring them with a high level of accuracy, such as in the work of Checchin *et al.* [6] using a 3D niobium cavity to isolate the silicon substrate and quantify its loss contribution. In another study, Romanenko *et al.* [7] used the Nb film by itself as a three-dimensional (3D) resonator and showed that two-level system (TLS) defects present in the niobium native oxide dominate the losses in 3D resonators below 1.4 K and at low fields in a similar manner to what has been observed in 2D superconducting qubits [8,9]. In this sense, niobium cavities can be seen as a tool to investigate the microscopic origins of TLS defects in Nb-based quantum devices and a platform on which to experiment different approaches to suppress these loss channels in niobium 2D qubits.

Upon exposure to air, niobium forms an amorphous oxide layer with varying stochiometries that is known to host TLS defects [7-12]. Recent studies on three-dimensional niobium resonators have shown that these native oxide layers contribute substantially to the degradation of quality factors (Q), reducing it to about 2 × $10^{10}$ in 1.3 GHz cavities [7, 8,10] in the TLS-dominated regime at low accelerating fields (≤ $10^{-2}$ MV/m). One way of increasing this quality factor has been demonstrated using high vacuum annealing at 340 °C for five hours, after which the cavity is maintained in a vacuum environment to prevent re-oxidation [7,10]. While effective, this method is impractical for quantum computing and sensing applications due to the need for sustained vacuum conditions. Another approach for suppressing two-level system losses has been reported by Kalboussi *et al.* [13] where the 3D resonators have been coated with a thin layer of aluminum oxide using atomic layer deposition (ALD), followed by a heat treatment at 650°C for 10 hours. This resulted in the reduction of the niobium native oxide while keeping the niobium metal passivated by the ALD layer. Even though this approach offers an air-stable improvement in the low field quality factor, it is worth noticing that the amorphous $Al_2O_3$ deposited by ALD itself exhibits (like all other dielectric materials) TLS losses that should inhibit improvement.

In this article, we report air and water-stable improvement of the low-field quality factor of a niobium single cell 1.3 GHz cavity after a thermal treatment at 650 °C for 10 hours in high vacuum (HV). We also achieve some of the best performances reported for 1.3 GHz niobium 3D resonators after air exposure and high pressure rinsing (HPR) in the TLS dominated regime with a $Q_0$ of 9 x $10^{10}$, at low fields ($\leq 10^{-2}$ MV/m) corresponding to a resonator lifetime τ ∼ 15s.

By performing X-ray photoelectron spectroscopy (XPS) and scanning transmission electron microscopy (STEM), we discovered that this thermal treatment changes the chemical and structural nature of the niobium oxides after re-exposition to air and high pressure rinsing (HPR). In particular, we reveal the appearance of crystalline regions in the native oxide layer, which is associated with low TLS contributions and explains the high quality factor measured.

## II. EXPERIMENTAL DETAILS

In this study, we report RF results from a 1.3 GHz niobium cavity before and after annealing at 650 °C for 10 hours in high vacuum (pressure < $10^{-6}$ mbar). Before the thermal treatment, the cavity underwent a standard electro-polishing process to prepare its surface [14] followed by a high pressure rinsing with ultra-high purity water under a pressure of 90 bar for 1 1/2 hour [15]. Subsequently, the cavity was mounted for cryogenic testing in an ISO5 clean room environment. The cryogenic testing was performed in the Synergium vertical test facility at CEA in which the cavity was submerged in a dewar of liquid helium then cooled through pumping to 1.4 K. RF test was conducted by employing a phase-lock loop to lock the cavity onto its resonance frequency, enabling the derivation of the intrinsic quality factor ($Q_0$) and field. The low-field region quality factor was measured using the cavity ring-down method after the RF power was turned off, as described in [16]. After the test, the niobium cavity was introduced to the vacuum oven and underwent an annealing at 650°C for 10 hours in a pressure lower than $5.10^{-6}$ mbar using a ramp of 6 °C/min. After cooling down passively, the vacuum was broken in the oven and the cavity was transported to the clean room again and remained air-exposed there for few weeks before undergoing a second high pressure rinsing and preparation for the RF test.

In order to investigate the chemical, structural and electronic properties changes on the niobium surface, X-ray photoelectron spectroscopy (XPS), transmission electron microscopy (TEM) and point contact tunneling spectroscopy (PCT) analyses were performed on two types of samples from air-exposed cavity-grade niobium coupons : (1) electro-polished then high-pressure rinsed and (2) electro-polished, high-pressure rinsed then annealed in the same conditions used for the cavity followed by a second HPR.

XPS measurements were acquired using an Escalab 250 XI spectrometer (Thermo Fisher Scientific, Waltham, MA, USA) equipped with a monochromatic X-ray Al-Kα source (hν = 1486.6 eV) and charge compensation system. The diameter of the analytical spot size is 900 µm. The binding energies were calibrated against the C1s binding energy set at 284.8 eV. The spectra were treated by means of CasaXPS software [17]

For to the STEM analysis, cross section lamellae were prepared in a FIB microscope using a Ga ion source at University of Lille. Note that for the FIB preparation, the sample surface was firstly protected by a deposition of carbon followed by another layer of platinum and a standard lift-out procedure was performed. Bright field and high-angle annular dark-field (HAADF) images of the cross-section samples were acquired at LPS using a Nion Ultrastem 200 scanning transmission electron microscope operating at 100 kV. The probe convergence angle was 35 mrad.

The tunneling spectroscopy was performed on a home-made apparatus at a temperature of 1.8 K as described in [18].

*Contact authors: Yasmine.kalboussi@cea.fr
Thomas.proslier@cea.fr

## III. RESULTS AND DISCUSSION

### A. Resonator RF tests

Fig. 1 shows the results of the RF tests performed on the 1.3 GHz niobium cavity at 1.43 K. The baseline RF test (green curve) shows the typical saturation of the quality factor $Q_0$ of air-exposed electro-polished niobium cavities without any thermal treatments, with values at low fields around $1 \times 10^{10}$ at 0.01 MV/m as has been reported [8,10,13] and which is characteristic of the presence of two-level systems. The blue curve represents the RF test after annealing the same niobium cavity at 650°C for 10 hours at high vacuum and exposing it to air for few weeks and high pressure rinsing. After this treatment, the $Q_0$ saturation level increased approximately tenfold, attaining an unprecedented value of $9 \times 10^{10}$ at saturation (0.001 MV/m) after air-exposure and high pressure rinsing.

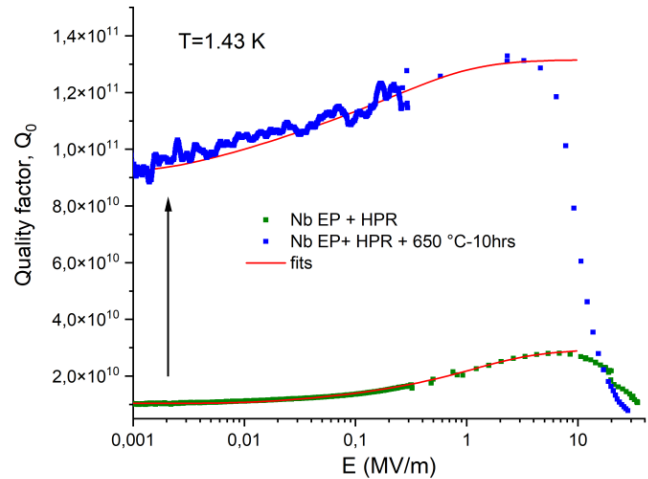

FIG 1. $Q_0$ versus E results of 1.3 GHz niobium cavity before and after annealing at 650 °C-10hrs

It is worth mentioning that, among reported RF tests on bulk niobium cavities [7,8,10,13], this 10-hour treatment at 650° C showed one of the highest performances in the TLS-dominated regime. As a matter of fact, previous studies tested Nb cavities with different treatments and in particular annealing at 340 °C for 5 hours which showed improvements over the electro-polished baseline with a quality factor value of $7 \times 10^{10}$ at 0.01 MV/m [7,10]. This improvement was ascribed to the complete dissolution of the native niobium pentoxide $Nb_2O_5$ during this annealing. However, one needs to keep the cavity under vacuum to prevent $Nb_2O_5$ from re-growing upon exposure to air; otherwise, the cavity re-exhibits EP performances. The fact that annealing the niobium cavity at 650 °C for 10 hours results in significantly higher $Q_0$ that is maintained after HPR and exposure to air suggests that time-stable and irreversible modification has occurred at the niobium surface of the resonator.

The red lines in Fig.1 are fits using the interacting and non-interacting two level system (TLS) model described in [19]. The fitting and the extracted TLS parameters are shown in Table I. The fit parameters for the baseline EP are consistent with the ones obtained in [13,19] for a niobium EP cavity with similar RF performances. The c parameter that describes the saturating value of the Q at very low electrical fields is more than 10 times lower after the annealing. Following ref [19], the density of TLS defect, $\sigma_{TLS}$, and the dielectric losses, $\tan(\delta_{TLS})$, can be calculated from the fitting parameters assuming a dielectric constant for $Nb_2O_5$ of 30 and knowing the thickness of the oxides as described later. These fitting values reveal an order of magnitude reduction in TLS defect density and losses in the annealed Nb cavity as compared to native niobium oxides present in the cavity baselines.

In order to identify the microscopic origins of these changes in the RF performances and TLS losses, we performed XPS measurements on niobium samples that underwent the same processes as the cavity, along with high-angle annular dark-field imaging (HAADF) and Fast Fourier Transform (FFT) analysis.

*Contact authors: Yasmine.kalboussi@cea.fr
                  Thomas.proslier@cea.fr

TABLE I. Fitting parameters using the interacting and non-interacting TLS model [19].

| Treatments | $c$ $(C^2/J)$ | $E_C$ $(V/m)$ | $\xi$ | $1/Q_{non-TLS}$ | $\sigma_{TLS}$ $(cm^{-2})$ | $tan(\delta_{TLS})$ |
|---|---|---|---|---|---|---|
| Baseline EP | $1.6 \times 10^{-23}$ | $1.4 \times 10^4$ | 200 | $3.4 \times 10^{-11}$ | $3.7 \times 10^{11}$ | $1.7 \times 10^{-3}$ |
| EP+650 °C-10 hrs | $8.5 \times 10^{-25}$ | $2 \times 10^3$ | 600 | $0.76 \times 10^{-11}$ | $1.9 \times 10^{10}$ | $1.5 \times 10^{-4}$ |

## B. Chemical and structural analysis of the Nb surface

Fig. 2 summarizes the surface analysis obtained on a cavity-grade electro-polished Nb sample that underwent a high pressure rinsing without any thermal post-treatment. XPS provides a detailed analysis of the Nb oxidation state and chemical environment. High-resolution spectra of the Nb 3d core levels were recorded using a constant pass energy of 20 eV. In order to fit the peaks, we used an asymmetrical line shape for the metallic component of the Nb 3d core levels and symmetrical mixed Gaussian-Lorentzian line shapes to fit the other chemical states. An area constraint of 3:2 was applied to the spin-orbit doublets of Nb $3d_{3/2}$ and Nb $3d_{5/2}$ levels and their binding energies were set apart by 2.7 eV. Finally, the valence state of every peak was identified by its binding energy from the literature [20-21]. The binding energy ranges for Nb $3d_{5/2}$ in different valence states are as follows: $202.2 \pm 0.05$ eV, $203.1 \pm 0.2$ eV, $204.3 \pm 0.5$ eV, $205.5 \pm 0.2$ eV and $207.6 \pm 0.1$ eV which are respectively considered as Nb, $Nb_2O$, NbO, $NbO_2$ and $Nb_2O_5$. The results of peak fitting for Nb 3d chemical states is shown in Fig. 2(a). We observe that the surface is composed mostly of $Nb_2O_5$ (86.6%), a small fraction of sub-oxides such as $Nb_2O$ (1%), NbO (1.6%) and $NbO_2$ (2.7%) and metallic niobium (8%). This oxide composition is typical of an un-coated electro-polished Nb sample subject to a high pressure rinsing with a native oxide composition dominated by $Nb_2O_5$, in agreement with previous studies [13,20-21].

The HAADF image in Fig. 2(b) shows a uniform oxide layer of 6.5 nm on top of the crystalline niobium substrate. Local fast Fourier transform (FFT) analysis confirmed that the native oxide layer is amorphous while the niobium substrate is crystalline, with a cubic (Im-3m) lattice. The average thickness of niobium's native oxide after exposure to air is typically around $5 \pm 1.5$ nm [22, 23], although this varies significantly depending on the surface preparation and oxidation conditions. In our case, it is not surprising that the native oxide is thicker, as high pressure rinsing causes a notable increase in the oxide layer thickness [22, 23].

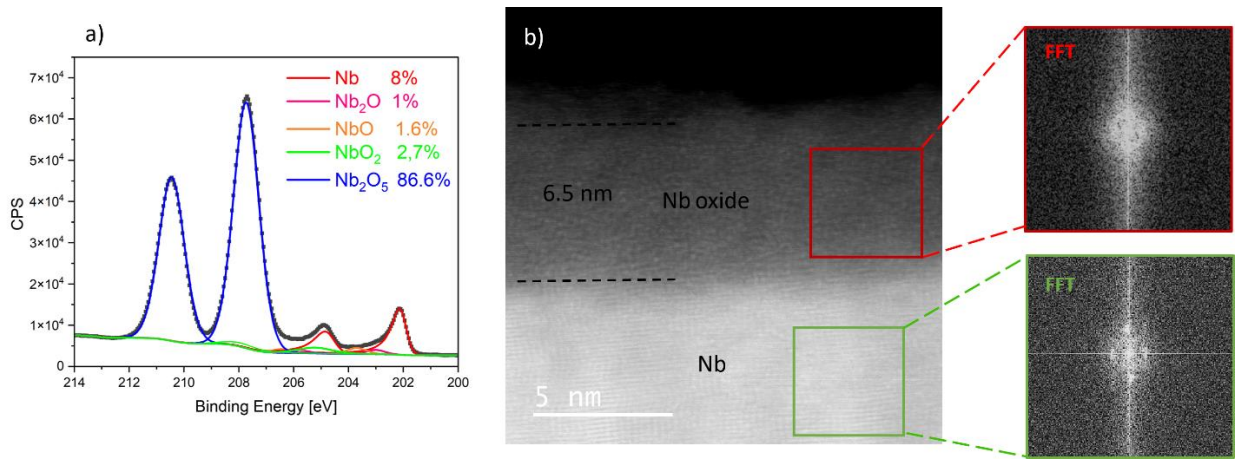

*Contact authors: Yasmine.kalboussi@cea.fr
Thomas.proslier@cea.fr

**FIG 2. (a) XPS spectrum of Nb-3d core levels of electro-polished niobium+ HPR (b) HAADF imaging and local FFT analysis on electro-polished Nb + HPR**

After annealing at 650°C for 10 hours followed by HPR and air exposure, XPS reveals an altered composition of the surface. Fig. 3(a) shows a decrease in $Nb_2O_5$ composition from 86.6% to 61.1% along with an increase in the percentage of the metallic niobium and sub-oxides. In particular, we witness the emergence of clear peaks corresponding to $Nb_2O$ and $NbO$ with contributions of 12% and 8.5 % to the Nb 3d signal respectively. The HAADF imaging shown in Fig. 3(b) reveals a uniform and a significant reduction in the oxide layer thickness from 6.5 nm to 4 nm.

The oxide layer alteration is the result of the furnace baking under HV and re-oxidation after air exposure and high pressure rinsing. Based on previous heat-treatment studies [10,13,21], niobium pentoxide gets completely dissolved when annealed at temperature higher than 340 °C and gives rise to other sub-oxides, mainly NbO. When exposed to air, the formation of NbO on the surface of Nb significantly raises the potential barrier for oxygen uptake from the air, thereby slowing the oxide's growth. While studying niobium heat treatments in the range of 250 °C-800 °C over a few hours, Yu *et al.* [21] found that the proportion of $Nb_2O_5$ progressively decreases after air exposure as the heat-treatment temperature rises within the range of 300–600 °C. After 600 °C heat treatment, their samples show the lowest concentration of $Nb_2O_5$ and the largest concentration of NbO along with the emergence of peaks of $Nb_2O$. These results are in perfect agreement with our observations: it seems that our annealing at 650 °C for 10 hours results in such a formation of an NbO- and $Nb_2O$-rich layer that inhibits significantly the formation of a thick $Nb_2O_5$ layer formation, even after the high pressure rinsing step which is known to favor its growth. This can explain the significant difference in TLS losses at low RF fields and the associated decrease in $\sigma_{TLS}$ and $\tan(\delta_{TLS})$ parameters extracted from the fits. In contrast, annealing at lower temperatures such as 350 °C [21] does not seem to create enough NbO to slow the re-growth of $Nb_2O_5$ after exposure to air. This could explain why the reduction of the TLS observed by Romanenko *et al.* [7,10] after annealing at 340 °C does not persist upon exposure to air.

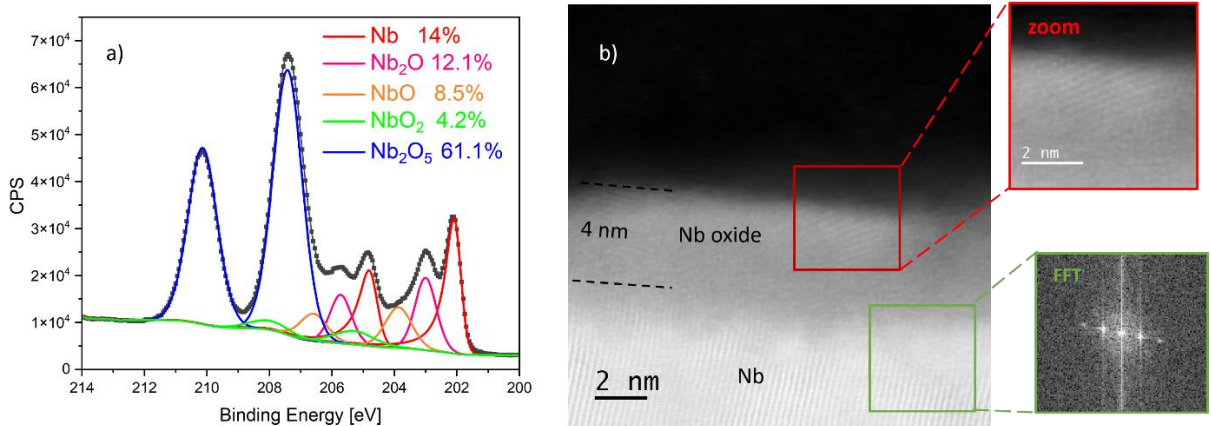

**FIG 3. (a) XPS spectrum of Nb-3d core levels of annealed niobium at 650 °C-10 hours + HPR (b) HAADF imaging and local FFT analysis on annealed niobium at 650 °C-10 hours Nb+HPR**

More interestingly, we notice the formation of clear crystalline regions in the oxide layer. Unlike the native oxide on top of the un-annealed sample, which is thick and amorphous, the annealing at 650 °C during 10 hours results in the formation of nano-scale crystallites inside the overall amorphous matrix of the oxide. Fig. 4 shows different HAADF images of the Nb sample annealed at 650 °C for 10 hours and high pressure rinsed, along with local FFT analysis. It can be seen that in some regions such as zone A, the crystallite is located near the interface with the metallic niobium. FFT analysis in this region shows a d-spacing of 2.1 Å which can be associated with NbO (200) planes [24]. In zones

*Contact authors: Yasmine.kalboussi@cea.fr
Thomas.proslier@cea.fr

B and C, we observe another configuration where the crystallite is formed at the surface on top of an amorphous layer of oxide. FFT analysis in these zones shows d-spacings of 1.5 Å, 2.4 Å and 2.9 Å which can be associated respectively with the (220), (111) and (110) planes of NbO [24]. In other regions such as Zone D, the crystal grows starting from the interface with the metal and continues up to the surface. In this zone, we measure a d-spacing of 2.4 Å which can be associated with the (111) planes of NbO [24]. Based on these observations, a plausible scenario is that during the 10-hour annealing at 650°C the original amorphous $Nb_2O_5$ decomposes to a nanocrystalline layer of NbO and possibly $Nb_2O$ ( although there is no reported crystalline structure of $Nb_2O$ in literature). Upon exposure to air and high pressure rinsing, $Nb_2O_5$ regrows surrounding the crystallites of NbO and sometimes underneath it (zone B and C). In other cases, the crystalline NbO layer formed during the annealing is already passivating and thus no $Nb_2O_5$ is formed after the exposure to air. Similar observations have been reported by Hellwig *et al.* [25] who witnessed the formation of a crystalline and passivating layer of NbO (111) on top of a Nb (110) film. Our discovery of these nano-crystalline regions is in perfect agreement with the surprisingly low TLS losses of the Nb 1.3 GHz cavity.

Shortly after the discovery that qubit energy relaxation and decoherence were linked to TLS in amorphous materials, several efforts were made to create fully crystalline tunnel-junction barriers. Despite the technological difficulties of this approach, Seongshik *et al.* [26] showed the benefit of integrating single crystal epitaxial $Al_2O_3$ tunnel barriers into Josephson phase qubits and presented measurements showing a correlation between the crystallinity of the tunnel barrier and the density of TLSs in the qubit. In another study, Patel *et al.* [27] incorporated a single crystal silicon shunt capacitor into a Josephson phase qubit and showed that the superior dielectric loss of the crystalline silicon leads to a more than doubling of the qubit energy relaxation times compared to those seen in amorphous phase qubits. Our findings confirm the dominant participation of the niobium pentoxide $Nb_2O_5$ in TLS losses of Nb-based quantum devices and provides evidence that crystalline NbO can be a much less dissipative alternative for passivating Nb-based devices.

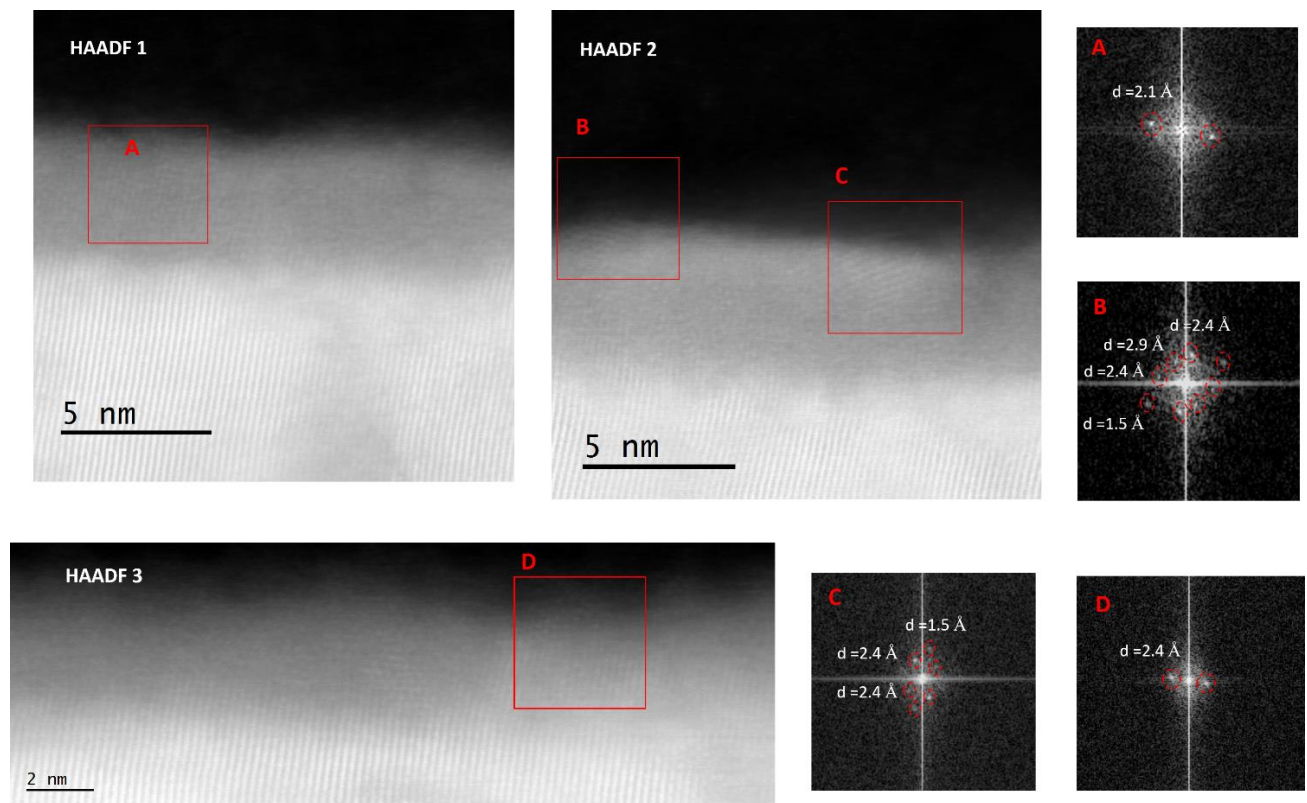

*Contact authors: Yasmine.kalboussi@cea.fr
Thomas.proslier@cea.fr

**FIG 4. HAADF imaging on three different locations on the annealed niobium samples at 650 °C-10 hours + HPR and corresponding local FFT analysis**

## C. Tunneling spectroscopy

We measured the surface superconducting properties by means of point contact tunneling (PCT) spectroscopy as described in [18]. In our setup, the junctions were formed by approaching the sample surface with an Au tip, creating a superconductor-insulator-normal (SIN) junction where the insulator is the oxide layer on the sample surface. The PCT data were analyzed using the expression for the differential tunneling conductance:

$$\frac{dI}{dV} \propto \int N_S(E) \left[ -\frac{\partial f(E + eV)}{\partial (eV)} \right] dE \tag{1}$$

where $I$ is the current flowing through the SIN junction under a difference of potential $V$, $N_S(E)$ is the superconducting density of states (DOS), and $f(E)$ is the Fermi function.

We considered two models to describe $N_S(E)$. The first is the widely known Dynes model given by the formula [28]

$$N_S(E) = N_S \mathrm{Re} \left[ \frac{E + i\Gamma}{\sqrt{(E + i\Gamma)^2 - \Delta^2}} \right] \tag{2}$$

where $N_S$ is the DOS at the Fermi surface in the normal state, $\Delta$ is the superconducting gap and $\Gamma$ is the phenomenological quasiparticle lifetime broadening parameter.

The second model is based on the Usadel equations for a proximity-coupled dirty thin normal layer (N) on a surface of a bulk superconductor (S) [29]. The resulting DOS is controlled by fours parameters: the Dynes broadening parameter $\Gamma$, the bulk pair potential $\Delta$, and two dimensionless parameters $\alpha$ and $\beta$, which are determined by the N-layer thickness and the N-S interface transparency:

$$\alpha = \frac{d}{\xi_S} \frac{N_N}{N_S}, \ \beta = \frac{4e}{\hbar} R_B N_N \Delta d. \tag{3}$$

Here $\xi_S$ is the bulk coherence length, $d$ is the N-layer thickness, $N_N$ and $N_S$ are the DOS at the Fermi surface in the normal state of the N-layer and S-layer, respectively, and $R_B$ is the contact resistance of the N-S interface.

We performed PCT measurements on the same samples used for the TEM cross sections measurements as described previously. For the reference sample that received an EP and an HPR, we could not gather enough statistic because the tunnel junction resistances were too high (> 2-5 GΩ) to be measured for our experimental PCT set up. Such high tunnel junction resistances is most likely due to the thick native niobium oxide layer (6.5-7 nm) measured by TEM in Fig. 2. We therefore decided to use as a reference sample an electropolished Nb coupon rinsed in water but without the HPR step. As far as we know, HPR main effect is to increase the surface oxide thickness, so it is reasonable to assume that it does not significantly alter the DOS. For each sample, about ∼100 junctions were measured over an area of ∼100 μm x 100 μm at T=1.8 K.

Typical tunneling conductance spectra for both samples with the corresponding fits are shown in Fig. 5. For the Nb+HPR+ 650 °C-10 h + HPR sample (Fig. 5 (a)),the fits obtained from the Usadel proximity model are in very good agreement with the data, whereas the fits extracted from the Dynes model fail to accurately reproduce the spectra. This indicates the presence of a normal metal layer on the niobium surface, in agreement with the TEM observations in section B that reveal the presence of metallic NbO nanocrystals. In contrast, the Dynes model fit well the reference sample spectra shown In Fig. 5 (b) with no indication of a proximity effect, which is also consistent with the TEM cross sections of the referenced sample shown in section B that show an abrupt transition from an amorphous $Nb_2O_5$ to the metallic Nb. In the Usadel theory, the gap induced in the normal metal layer, referred as $\varepsilon_0$ [30] and sometime called the minigap, is given by the equation:

*Contact authors: Yasmine.kalboussi@cea.fr
Thomas.proslier@cea.fr

$$\beta = \frac{1}{(\varepsilon_0/\Delta)} \left(\frac{1-(\varepsilon_0/\Delta)}{1+(\varepsilon_0/\Delta)}\right)^{1/2} \qquad (4)$$

The effective superconducting gap $\varepsilon_0$ is the one measured on the conductance spectra in Fig. 5(a) and approximately given by half the energy separation between the quasiparticle peaks. The gap $\Delta$ in the superconductor underneath the normal metal layer is inferred from the fits but not directly probed by tunneling spectroscopy. In the Dynes model, there is only one superconducting gap $\Delta$.

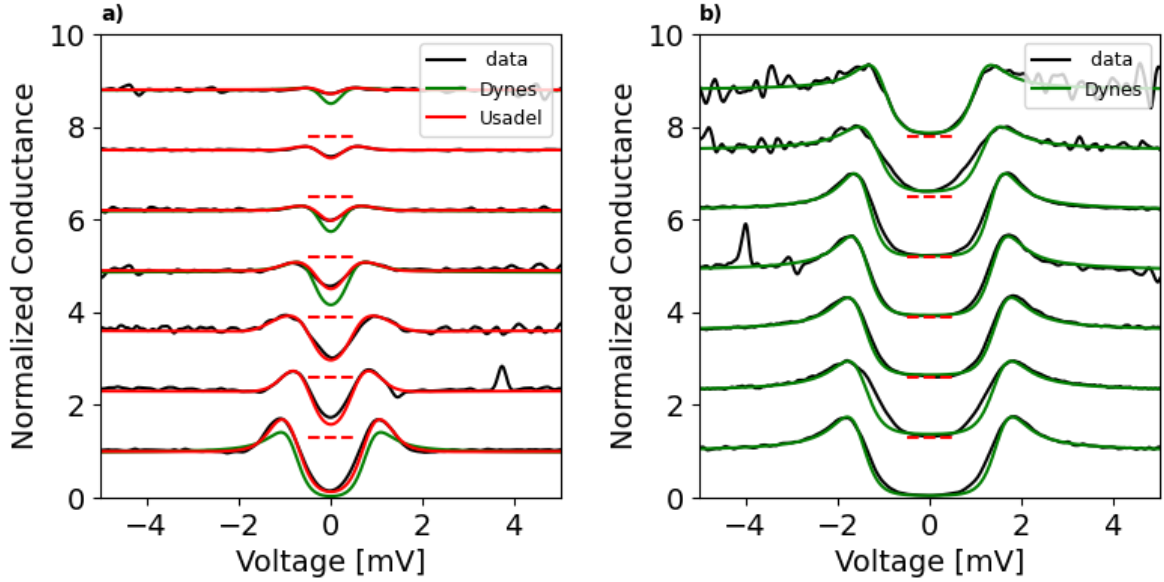

**FIG. 5. a) Selected tunneling conductance curves from Nb+HPR+650°C-10 h+HPR sample measured at T=1.8 K and the corresponding fits using Usadel and Dynes models. The parameters $\Delta$ ,$\Gamma$, $\alpha$ , $\beta$, extracted from the proximity fits are from top to bottom: 0.64, 0.09, 0.094, 3.18 \ 0.8, 0.07, 0.15, 2.9 \ 1.01, 0.09, 0.2, 3.1\ 1.13, 0.1, 0.2, 2\ 1.35, 0.09, 0.15, 1.44\ 1.05, 0.045, 0.13, 0.93\ 1.28, 0.045, 0.15, 0.47. The parameters $\Delta$, $\Gamma$, extracted from Dynes fits are from top to bottom: 0.19, 0.001\ 0.32, 0.02\ 0.46, 0.01 \ 0.83, 0.001. b) Selected tunneling conductance from Nb+EP sample measured at T=1.8 K and the corresponding fits using Dynes model. The parameters $\Delta$, $\Gamma$, extracted from Dynes fits are from top to bottom: 1.08, 0.06\ 1.26, 0.12\ 1.42,0.02\ 1.47, 0.05\ 1.51, 0.1\1.55, 0.07\1.53, 0.07.**

The statistic of the superconducting gaps $\Delta$ extracted from the fits on the Dynes and Usadel models and the corresponding cartographies are represented in Fig. 6.

The histogram (Fig. 6 (a)) shows for the annealed sample a large distribution of $\Delta$ values centered around $0.9 \pm 0.2$ meV, significantly lower than the well-established bulk Nb gap value around 1.5 meV [30]. In contrast, the histogram distribution for the reference sample is peaked around the bulk Nb gap value of 1.45 meV with a smaller spread of $\pm$ 0.17 meV. The cartography represented in Fig. 6 (b) and (c) illustrates these differences with a significant inhomogeneity observed across different locations for the annealed sample, with some areas showing a strong suppression of $\Delta$, reaching values as low as ~0.4 meV. The low values of superconducting gap $\Delta$ for the annealed sample indicates that the superconductor underneath the normal metal layer is systematically strongly reduced as compared to the bulk Nb gap.

*Contact authors: Yasmine.kalboussi@cea.fr
Thomas.proslier@cea.fr

The histograms for the others parameters extracted from Usadel fits of the annealed sample spectra, namely the minigap $\varepsilon_0$, $\beta$ and $\alpha$ and their associated spatial variations are represented in Fig. 7. It is notable that the interface transparency $\beta$ varies significantly between a nearly transparent N-S interface ($\beta \ll 1$) to a weak N-S coupling ($\beta \gg 1$) [29]. These variations can be explained by considering that the transparency of the N-S interface strongly depends on the proximity of the NbO nanocrystals to the Nb. For crystals directly above the Nb, as observed in regions A and D of Fig. 4, we expect a better interface contact and, therefore, a small $\beta$. On the other hand, for crystals formed at the surface on top of the amorphous layer of oxide, as seen in regions B and C of Fig. 4, we would expect weak coupling since the N and S layers are separated by an insulating barrier, which increases the contact resistance $R_B$.

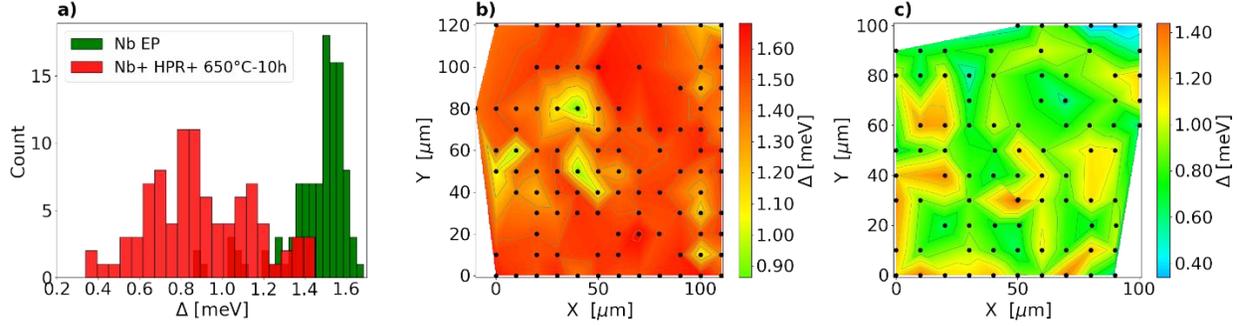

**FIG. 6. (a): Histograms of the Dynes parameters Δ extracted from the fits on the measured conductance curves of Nb+HPR+650 °C-10 h+HPR (green) and Nb+EP (red) samples over an area of ~100 μm x 100 μm at T = 1.8 K. Maps of the superconducting gap Δ extracted from of the fits of the reference sample (b) and annealed sample (c). The black points represent the location of the tunnel junctions.**

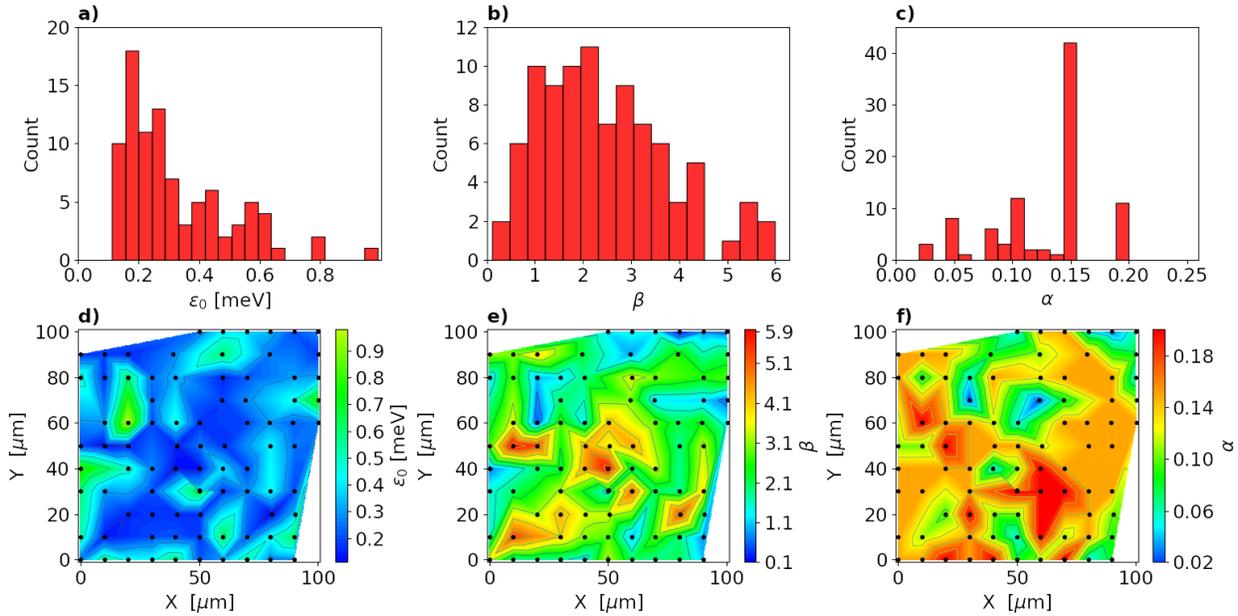

**FIG. 7. Histograms (top) and maps (bottom) of $\varepsilon_0, \alpha, \beta$ extracted from the Usadel fits on the measured conductance curves of Nb+HPR+650 °C-10 h+ HPR sample over an area of ~100 μm x 100 μm at T = 1.8 K. The black points represent the location of the tunnel junctions.**

*Contact authors: Yasmine.kalboussi@cea.fr
Thomas.proslier@cea.fr

Despite the complexity of the nanocrystals distribution within the amorphous oxide matrix, we can perform a simple calculation to estimate the thickness of the N-layer on top of the Nb. From $\alpha$ statistics, we obtained a peaked value at $\alpha = 0.15 \pm 0.05$. Assuming $N_N \sim N_S$ in Eq. (3), we have $\alpha \sim d / \xi_S$. If we take $\xi_S \sim 40$ nm for bulk Nb, the estimated thickness of the N-layer on top of the Nb is $d \sim 5$ nm, which is close to the $\sim 4$ nm observed in the TEM images for the nano crystallites.

## V. CONCLUSIONS

In conclusion, we have investigated the effect of high temperature annealing on the RF performances of Nb 1.3 GHz resonators in the TLS dominated regime at RF field intensities below $10^{-2}$ MV/m. An annealing in vacuum at 650 °C for 10 hours resulted in a tenfold enhancement of the low-field quality factor after air exposure and high pressure rinsing, which is caused by an alteration of the chemical composition of the niobium native oxide and its partial crystallization as probed by STEM, XPS and PCT. Our findings provide a new route for suppressing TLS defects in superconducting Nb-based resonators, which consists in promoting the niobium oxide crystallization, instead of uniquely dissolving the $Nb_2O_5$. Our findings on the chemical composition and structural evolution mechanism of Nb surface oxides can also be used to guide the thermal treatments of other superconducting quantum devices.

## ACKNOWLEGEMENT

This project has received funding from the region Ile de France project SESAME AXESRF, Internal CEA funds through the Program Transverse de Compétence- Matériaux et Procédés, the European Union's Horizon 2020 Research and Innovation program under Grant agreement No 101004730 and Grant agreement No 730871.

## AUTHORS DECLARATIONS

**Conflict of Interest:** Authors Y.K and T.P. has Patent N° PCT/FR2023/051937 pending.

**Authors Contributions: Yasmine Kalboussi:** Conceptualization (equal); Data curation (equal); Formal analysis (equal); Investigation (lead); Methodology (equal); Resources (equal); Validation (equal); Visualization (equal); Writing–original draft (lead). **Ivana Curci**: Investigation (equal); Formal analysis (equal); Data Curation (equal); Methodology (equal); Resources (equal); Validation (equal); Visualization (equal); Writing-original draft (equal); **David Troadec**: Resources (equal). **Frederic Miserque**: Investigation (equal). **Nathalie Brun** : Investigation (equal); Formal analysis (equal); Data Curation (equal). **Michael Walls**: Investigation (equal); Formal analysis (equal); Data Curation (equal); Writing-Reviewing and Editing (equal). **Gregoire Julien:** Resources (equal). **Fabien Eozenou:** Resources (equal). **Matthieu Baudrier:** Investigation (equal). **Luc Maurice:** Investigation (equal). **Quentin Bertrand:** Resources (equal). **Patrick Sahuquet:** Resources (equal). **Thomas Proslier:** Conceptualization (equal);

*Contact authors: Yasmine.kalboussi@cea.fr
Thomas.proslier@cea.fr

Data curation (equal); Formal analysis (equal); Funding Acquisition (lead); Supervision (lead); Investigation(equal); Methodology(equal); Validation (equal); Visualization (equal); Writing–original draft (equal).

---

[1] H. Paik, D. I. Schuster, L. S. Bishop, G. Kirchmair, G. Catelani, A. P. Sears, B. R. Johnson, M. J. Reagor, L. Frunzio, L. I. Glazman, S. M. Girvin, M. H. Devoret, and R. J. Schoelkopf, Phys. Rev. Lett. 107, 240501 (2011)

[2] H. Paik, A. Mezzacapo, M. Sandberg, D. T. McClure, B. Abdo, A. D. Córcoles, O. Dial, D. F. Bogorin, B. L. T. Plourde, M. Steffen, A. W. Cross, J. M. Gambetta, and J. M. Chow, Phys. Rev. Lett. 117, 250502 (2016).

[3] M. Reagor, W. Pfaff, C. Axline, R. W. Heeres, N. Ofek, K. Sliwa, E. Holland, C. Wang, J. Blumoff, K. Chou, M. J. Hatridge, L. Frunzio, M. H. Devoret, L. Jiang, and R. J. Schoelkopf, Phys. Rev. B 94, 014506 (2016).

[4] Y. Wang, Z. Hu, B. C. Sanders and S. Kais, Qudits and High-Dimensional Quantum Computing, Front. Phys. 8 (2020) 1.

[5] Roy, Tanay, et al. *Qudit-based quantum computing with SRF cavities at Fermilab*. No. FERMILAB-CONF-24-0026-SQMS. Fermi National Accelerator Laboratory (FNAL), Batavia, IL (United States), 2024.

[6] Checchin, Mattia, et al. "Measurement of the low-temperature loss tangent of high-resistivity silicon using a high-Q superconducting resonator." *Physical Review Applied* 18.3 (2022): 034013.

[7] Romanenko, A., et al. "Three-dimensional superconducting resonators at T< 20 mK with photon lifetimes up to τ= 2 s." *Physical Review Applied* 13.3 (2020): 034032.

[8] Romanenko, A., and D. I. Schuster. "Understanding quality factor degradation in superconducting niobium cavities at low microwave field amplitudes." *Physical Review Letters* 119.26 (2017): 264801.

[9] Verjauw, J., et al. "Investigation of microwave loss induced by oxide regrowth in high-Q niobium resonators." *Physical Review Applied* 16.1 (2021): 014018.

[10] Bafia, D., et al. "Oxygen vacancies in niobium pentoxide as a source of two-level system losses in superconducting niobium." *Physical Review Applied* 22.2 (2024): 024035.

[11] Burnett, L. Faoro, and T. Lindström, "Analysis of high quality superconducting resonators: Consequences for TLS properties in amorphous oxides," Supercond. Sci. Technol. 29, 044008 (2016).

[12] A. Premkumar, C. Weiland, S. Hwang, B. Jäack, A. P. M. Place, I. Waluyo, A. Hunt, V. Bisogni, J. Pelliciari, A. Barbour, M. S. Miller, P. Russo, F. Camino, K. Kisslinger, X. Tong, M. S. Hybertsen, A. A. Houck, and I. Jarrige, "Microscopic relaxation channels in materials for superconducting qubits," Commun. Mater. 2, 72 (2021).

[13] Kalboussi, Yasmine, et al. "Reducing two-level systems dissipations in 3D superconducting niobium resonators by atomic layer deposition and high temperature heat treatment." *Applied Physics Letters* 124.13 (2024).

*Contact authors: Yasmine.kalboussi@cea.fr
Thomas.proslier@cea.fr

[14]   H. Diepers, O. Schmidt, H. Martens, and F. Sun, "A new method of electropolishing niobium," Phys. Lett. A 37, 139–140

[15]   P. Bernard, D. Bloess, W. Hartung, C. Hauviller, W. Weingarten, P. Bosland, and J. Martignac, "Superconducting niobium sputter-coated copper cavities at 1500 MHz," in Proceedings of 3rd European Particle Accelerator Conference (IEEE, 1991), pp.1269–1271

[16]   H. Padamsee, J. Knobloch, and T. Hays, RF Superconductivity for Accelerators (John Wiley & Sons, 2008).

[17]   N. Fairley, V. Fernandez, M. Richard-Plouet, C. Guillot-Deudon, J. Walton, E. Smith, D. Flahaut, M. Greiner, M. Biesinger, S. Tougaard, et al.,"Systematic and collaborative approach to problem solving using x-ray photoelectron spectroscopy," Applied Surface Science Advances 5, 100112 (2021).

[18]   N. Groll, M.J. Pellin, J.F. Zasadzinski and T. Proslier. "Point contact tunneling spectroscopy apparatus for large scale mapping of the surface superconducting properties". Rev. Sci. Instrum. 86, 095111 (2015).

[19]   N. Gorgichuk, T. Junginger, and R. de Sousa, "Modeling dielectric loss in superconducting resonators: Evidence for interacting atomic two-level systems at the Nb/oxide interface," Phys. Rev. Appl. 19, 024006 (2023).

[20]   Ma, Q., & Rosenberg, R. A. (2001, June). Surface study of niobium samples used in superconducting RF cavity production. In PACS2001. Proceedings of the 2001 Particle Accelerator Conference (Cat. No. 01CH37268) (Vol. 2, pp. 1050-1052). IEEE.

[21]   Yu, M., Pu, G., Xue, Y., Wang, S., Chen, S., Wang, Y., ... & Zhang, K. (2022). The oxidation behaviors of high-purity niobium for superconducting radio-frequency cavity application in vacuum heat treatment. Vacuum, 203, 111258.

[22]   Antoine, Claire. Materials and surface aspects in the development of SRF Niobium cavities. No. EuCARD-BOO-2012-001. 2012, page 64.

[23]   Tyagi, P. V., et al. "Study of HPR created oxide layers at Nb surfaces." Proc. LINAC 2012. 2012.

[24]   Diffraction pattern of NbO number 1010410 from the open database of crystallography

[25]   Hellwig, O., and H. Zabel. "Oxidation of Nb (110) thin films on a-plane sapphire substrates: an X-ray study." Physica B: Condensed Matter 283.1-3 (2000): 228-231.

[26]   Oh, Seongshik, et al. "Elimination of two level fluctuators in superconducting quantum bits by an epitaxial tunnel barrier." Physical Review B—Condensed Matter and Materials Physics 74.10 (2006): 100502.

[27]   Patel, U., et al. "Coherent Josephson phase qubit with a single crystal silicon capacitor." Applied Physics Letters 102.1 (2013).

[28]   Dynes, R. C., Narayanamurti, V., & Garno, J. P. (1978). Direct measurement of quasiparticle-lifetime broadening in a strong-coupled superconductor. Physical Review Letters, 41(21), 1509.

[29]   Gurevich, A., & Kubo, T. (2017). Surface impedance and optimum surface resistance of a superconductor with an imperfect surface. Physical Review B, 96(18), 184515.

[30]   T. Proslier, J. Zasadzinski, L. cooley, M. Pellin, J. Norem, J. Elam, C.Z. Antoine, R.A. Rimmer and P. Kneisel. IEEE transacations on applied superconductivity, 19, 3, 1404 (2009).

*Contact authors: Yasmine.kalboussi@cea.fr
              Thomas.proslier@cea.fr

\*Contact authors: Yasmine.kalboussi@cea.fr
Thomas.proslier@cea.fr